# Single- and Few-Layer WTe$_2$ and Their Suspended Nanostructures: Raman Signatures and Nanomechanical Resonances

Jaesung Lee[1†], Fan Ye[1†], Zenghui Wang[1], Rui Yang[1], Jin Hu[2],

Zhiqiang Mao[2], Jiang Wei[2], Philip X.-L. Feng[1*]

[1]*Department of Electrical Engineering & Computer Science, Case School of Engineering, Case Western Reserve University, 10900 Euclid Avenue, Cleveland, OH 44106, USA*

[2]*Department of Physics and Engineering Physics, Tulane University,
New Orleans, LA 70118, USA*

**Abstract**

**Single crystal tungsten ditelluride (WTe$_2$) has recently been discovered to exhibit non-saturating extreme magnetoresistance in bulk; it has also emerged as a new layered material from which atomic layer crystals can be extracted. While atomically thin WTe$_2$ is attractive for its unique properties, little study has been conducted on single- and few-layer WTe$_2$. Here we report the isolation of single- and few-layer WTe$_2$, as well as fabrication and characterization of the first WTe$_2$ suspended nanostructures. We have observed new Raman signatures of few-layer WTe$_2$ that have been theoretically predicted but not yet reported to date, in both on-substrate and suspended WTe$_2$ flakes. We have further probed the nanomechanical properties of suspended WTe$_2$ structures by measuring their flexural resonances, and obtain a Young's modulus of $E_Y \approx 80$GPa for the suspended WTe$_2$ flakes. This study paves the way for future investigations and utilization of the multiple new Raman fingerprints of single- and few-layer WTe$_2$, and for exploring mechanical control of WTe$_2$ atomic layers.**

---

[*]Corresponding Author. Email: philip.feng@case.edu.  [†]Equally contributed authors.

# Introduction

Following the advent of graphene [1], atomic layer two-dimensional (2D) crystals derived from layered materials, especially transition metal di-chalcogenides (TMDCs) such as $MoS_2$ [2], $WSe_2$ [3] and $MoTe_2$ [4], have generated enormous interests. The attractive new properties of these 2D crystals have spurred remarkable efforts on exploring new electrical [5], optical [6], and mechanical devices [7] based on these layered materials. Among various device structures, suspended, movable 2D flakes make a special platform with controllable mechanical degrees of freedom, which not only frees the crystal from the bonding of substrates, but can also enhance device performance and versatility. For examples, suspended graphene transistors exhibit ultra-high mobility up to 200,000 $cm^2V^{-1}s^{-1}$ [8], and suspended monolayer $MoS_2$ crystal can have orders of magnitude enhancement in photoluminescence (PL) intensity over its on-substrate counterpart [9]. Moreover, suspended structures based on atomic layers are essential for enabling 2D nanoelectromechanical systems (NEMS) such as ultrasensitive transducers and radio-frequency (RF) resonators. In particular, 2D materials possess remarkable mechanical properties such as high Young's modulus (*e.g.*, $E_Y \approx 1$TPa for graphene [10]) and ultra-high strain limits ($\sigma_{limit} \approx 30\%$ in black phosphorus [11]), making 2D materials highly promising for nanoscale sensors and actuators, and for integration with state-of-the-art NEMS to go across orders of magnitude length scales [12] and enable new integrated nanosystems.

Recent discovery of giant magnetoresistance in $WTe_2$ [13], resulting from the perfect balance between electron and hole populations [14], has stimulated great interests in $WTe_2$ as a new layered material with important potential device applications. Similar to studies on other forerunners of 2D crystals, Raman spectroscopy has been a very important means for characterization. While early Raman studies of $WTe_2$ (on $SiO_2$ substrates) have identified up to 7 Raman modes, many theoretically predicted modes remain yet to be observed [15,16,17,18,19]. In addition, suspended $WTe_2$ structures can offer better control over material properties in $WTe_2$ such as mechanically engineering the magnetoresistance through manipulating strain. Therefore, it is fundamentally important to systematically characterize the material properties including mechanical properties of suspended $WTe_2$ crystals.

In this work, we fabricate both on-substrate and suspended $WTe_2$ structures and investigate both the lattice vibration via Raman spectroscopy, and the coherent mechanical resonances of the entire suspended device structures. We observe a total of 12 Raman modes, all predicted by theoretical calculations [15,16,17,18,19], and their evolution over number of layers. We also extract $WTe_2$ crystal's mechanical properties such as the Young's modulus ($E_Y$) and pre-tension levels from nanomechanical resonance measurements.

# Raman Characterization of Single- and Few-Layer $WTe_2$

Figure 1a & 1b show the $WTe_2$ crystal structure. The atomic layers stack along the c-axis, and within each atomic layer the tungsten (W) atoms (lined up along the a-axis) are off-centered from their ideal sites (Fig. 1b), forming the distorted octahedral structure [20]: the W atoms are sandwiched between two Te atomic sheets; the three nearest Te atoms from each sheet form a triangular pyramid with the W atom, with the two resulting opposing pyramids rotated 180º (along the c-axis) from each other.

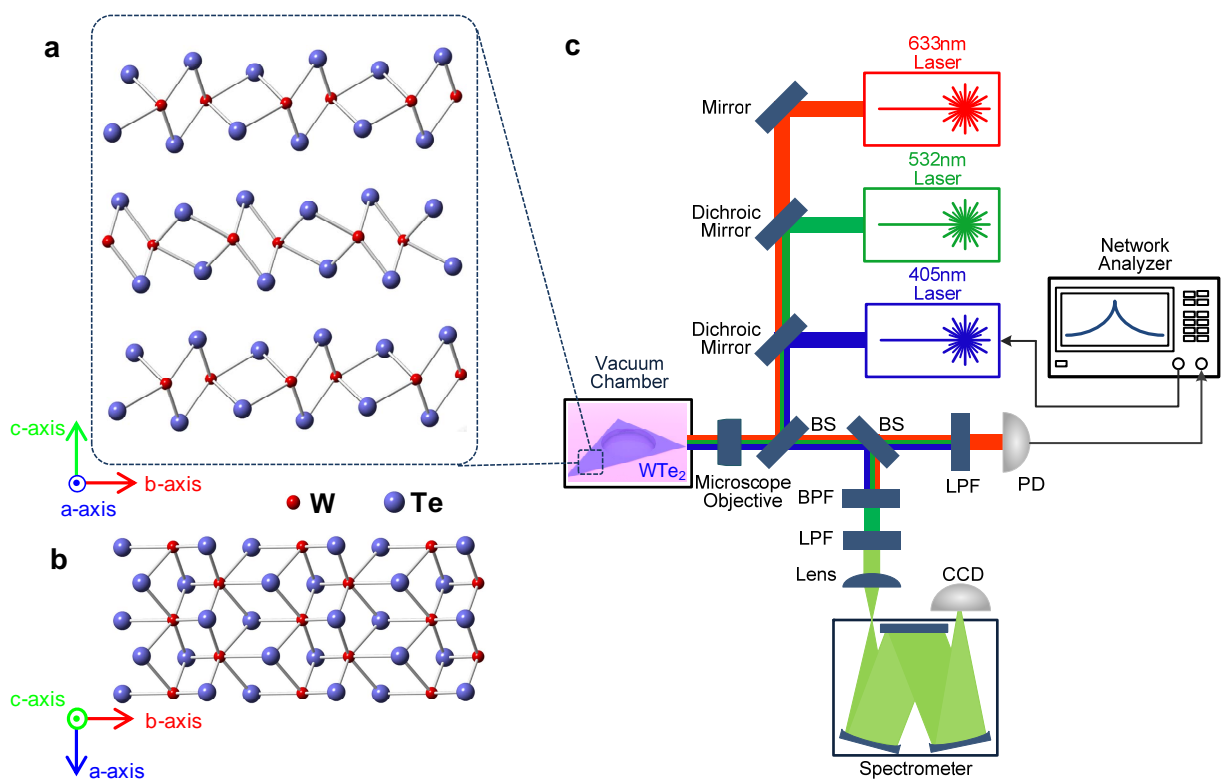

**Figure 1**: Crystal structure of $WTe_2$ and measurement system. (a) Side view and (b) top view of $WTe_2$ crystal structure (distorted octahedral). Red and blue spheres represent W and Te atoms respectively. (c) Combined Raman spectroscopy/interferometry measurement system. LPF, BPF, PD, and BS represent low-pass filter, band-pass filter, photodetector and beam splitter, respectively. All measurements are performed with samples in vacuum.

The Raman measurement system used in this study is integrated into a home-built optical interferometry for ultrasensitive detection of motions of the suspended samples, as illustrated in Fig. 1c (see Methods). Some of the representative samples are displayed in Fig. 2. We have carefully prepared and verified multiple single-layer (1L) flakes (Fig. 2a and more samples similar to this), and examined their Raman signatures, which repeatedly show 3 peaks in the 70–400 cm$^{-1}$ range (see detailed zoom-in plots in Fig. 2b). The 1L flakes exhibit clear features, which facilitates their identification reliably, through optical image contrast (as compared to other thicknesses, see Fig. 2a & 2c for examples), Raman spectra (clearly different than data from 2L and other few-layer flakes), and AFM imaging. All the 1L flakes are meticulously fabricated and promptly transferred into a vacuum chamber for timely and repeated measurements (see Methods). Figure 2c-g demonstrate optical images and AFM results from typical few-layer samples.

Figure 2h shows Raman spectrum measured from a six-layer (6L) $WTe_2$ flake with 12 Raman modes. To precisely determine the peaks' positions, we fit each peak with a Lorentzian curve, which gives Raman modes at 80.8, 86.9, 89.6, 110.4, 117.6, 120.8, 131.4, 134.8, 160.3, 163.9, 212.0, and 213.9 cm$^{-1}$ (referred to as P1 to P12, hereafter), as well as full width half maximum (FWHM) values of these peaks in the range of ~1.7 to 2.8 cm$^{-1}$. This is in clear contrast to previously reported Raman measurements in air, where only no more than 7 Raman modes with much broader FWHM (up to 10 cm$^{-1}$) [15,16,17,18,19]. These results confirm that comparatively high crystal quality of our WTe$_2$ flakes is preserved throughout measurements.

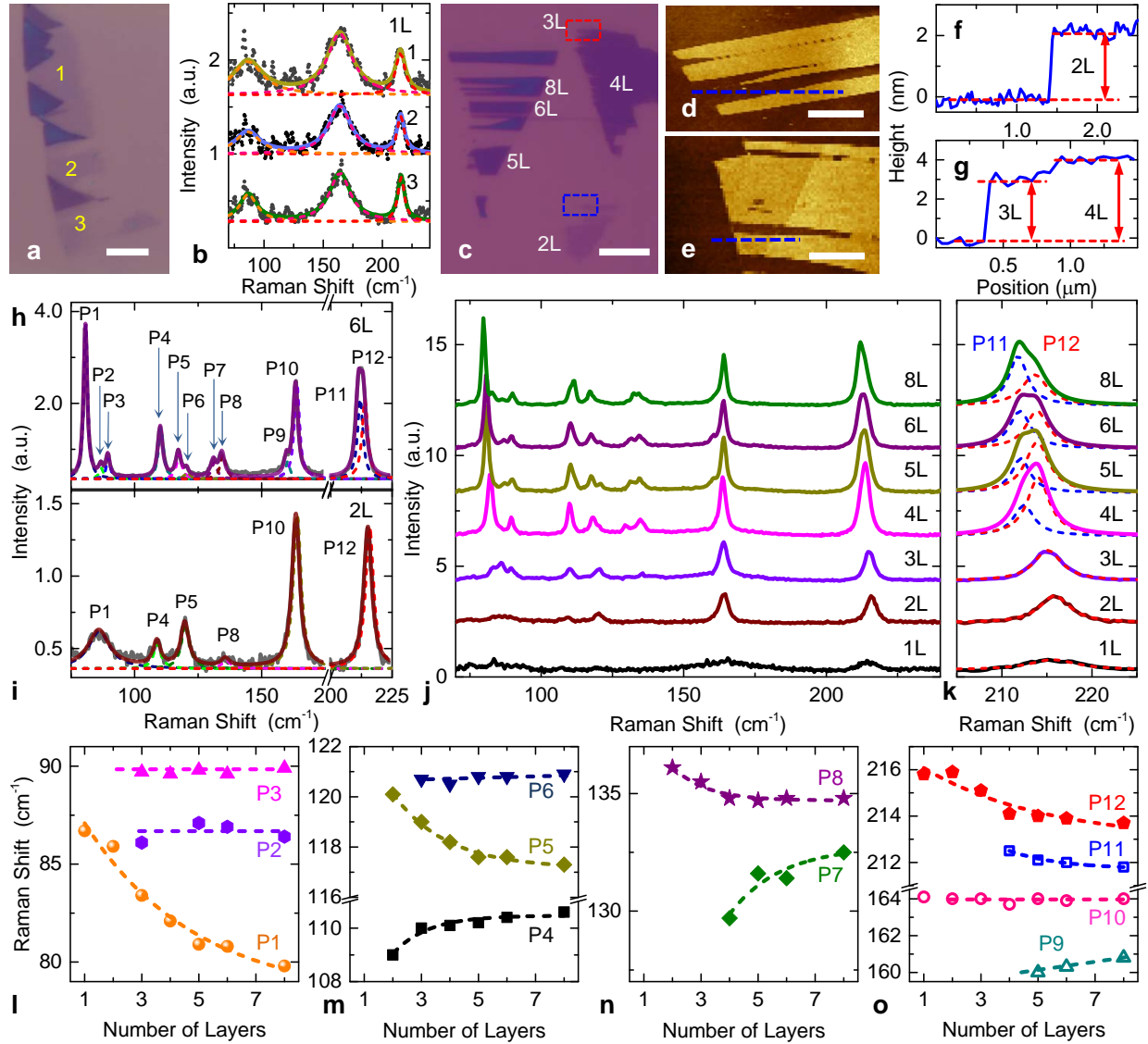

**Figure 2**: Raman spectroscopy of single-layer (1L) and few-layer WTe$_2$. (a) Optical image of representative 1L and few-layer WTe$_2$ flakes on 290nm SiO$_2$/Si substrate. Yellow numbers label the Raman measurement positions on 1L flakes. Scale bar: 5μm. (b) Measured Raman signal from 1L WTe$_2$ flakes. (c) Optical image of exfoliated WTe$_2$ flakes (number of layers labeled) on 290nm SiO$_2$/Si substrate. Scale bar: 5μm. (d)-(e) AFM images of areas outlined with (d) blue box and (e) red box in (c), respectively. Scale bars: 1μm. (f)-(g) AFM traces along dashed lines in (d) and (e), respectively. (h)-(i) Measured Raman spectra from (h) 6L and (i) 2L WTe$_2$, with Lorentzian fit. All peaks are labeled. (j) Evolution of Raman spectra from 8L to 1L. The data from 207cm$^{-1}$ to 222cm$^{-2}$ is zoomed in (k) to clearly show the evolution of P11 and P12 Raman modes. (i)-(o) Thickness dependence of Raman shift for all 12 modes.

We compare our measured Raman results with theoretical predictions [15,16,17,18,19], in which A$_1$, A$_2$, B$_1$, and B$_2$ Raman modes are expected for WTe$_2$. In our measurement, it is expected that only A$_1$ and A$_2$ Raman modes would appear since the laser polarization is mostly orthogonal to the crystalline c-axis. In particular, Raman modes at P1 (80.8 cm$^{-1}$), P10 (163.9 cm$^{-1}$), P11 (212.0 cm$^{-1}$), and P12 (213.9 cm$^{-1}$) exhibit high intensity among measured peaks and are well matched to the predicted Raman modes [15,16,17,18]. For some of $B_1$ and $B_2$ Raman modes that would not be favorably excited in our measurement scheme if considered only theoretically, we are still able to measure them due to out-of-plane polarization induced by the high numerical aperture (NA) of the microscope objective [21] in our experimental configuration.

By measuring crystals of different thicknesses, we explore the thickness dependence of Raman modes. Figure 2i shows that the Raman spectrum of 2L WTe$_2$ differs from that of multilayer flakes. We observe 6 clear peaks: P1 (86.1 cm$^{-1}$), P4 (109.0 cm$^{-1}$), P5 (120.1 cm$^{-1}$), P8 (136.1 cm$^{-1}$), P10 (164.0 cm$^{-1}$) and P12 (215.9 cm$^{-1}$). In addition, FWHM values range from 3.5 to 10 cm$^{-1}$, showing broadening compared with thicker samples. In the 1L WTe$_2$ flakes (Fig. 2a), the 3 Raman peaks exhibit larger FWHM values (7–22cm$^{-1}$) (see Fig. 2b, 2j, 2k) than in thicker layers. Figure 2j shows the evolution of Raman spectra from 8L down to 1L WTe$_2$. We observe that the intensities of P1 to P8 remain mostly unchanged from 8L down to 4L, but significantly decrease in 3L to 1L WTe$_2$. This thickness dependence of Raman intensities in P1 to P8 (mostly unchanged from 8L down to 4L, and then quickly decrease down in 3L and thinner WTe$_2$) may be explained by space group evolution from bulk ($C_{2v}$) to 1L ($C_{2h}$) WTe$_2$. Theoretical calculations predict that space group transition allows only a few Raman modes in 1L WTe$_2$ [16,18]; this may have contributed to significantly reducing the intensities of P1 to P8 modes, when the thickness is reduced down to 3L, 2L, and 1L. Similar space group evolution is also theoretically predicted in other TMDCs (*e.g.*, WSe$_2$, MoSe$_2$, WS$_2$, and MoS$_2$) from bulk to few-layer and monolayer structures [22].

Another important finding is the intensity evolution of P11 and P12 modes. Interestingly, Raman peak around ~215.9 cm$^{-1}$ is composed of two peaks (P11 and P12) with unique thickness dependence of Raman intensity (Fig. 2k). In this pair, P11 is very strong in 8L, and then gradually decline from 8L to 4L, and eventually too small to measure in 3L to 1L. This is similar to the behavior of P8. Meanwhile, P12 intensity increases as thickness decreases, and becomes dominant in 1L to 3L WTe$_2$. We attribute these observations also to the transition of space group in WTe$_2$ [16,18].

In addition, we observe clear and distinctive thickness dependence in Raman shift for all 12 peaks (Fig. 2l–o). The thickness dependence of the 12 modes exhibits three trends: softening (P4, P7, and P9), stiffening (P1, P5, P8, P11 and P12) and invariant (P2, P3, P6, and P10) as WTe$_2$ thickness decreases from 8L to 1L. Among all these Raman modes, we find that the P1 has the greatest shift ($\Delta\omega \approx 6$cm$^{-1}$), consistent with theoretical calculation [16]. This, together with the relative intensity and frequency change in P11 and P12, provides clear fingerprints for determination of number of layers in WTe$_2$ samples.

We further compare the Raman shift over sample thickness with theoretical predictions. 2H-MoS$_2$, the best-studied hexagonal TMDC, has two major Raman modes, $E^1_{2g}$ (in-plane) and $A^1_g$ (out-of-plane) [23]. $E^1_{2g}$ exhibits stiffening and $A^1_g$ shows softening as thickness decreases. Unlike MoS$_2$, theory predicts that Raman modes in WTe$_2$ are not confined to individual directions [16,19]: in particular, modes P6, P8 and P12 consist of both in-plane and out-of-plane motions, thus showing both stiffening and softening, as we have observed. Theory also predicts that mode P10 has displacement purely along the tungsten (W) atomic 1D chains, and is insensitive to crystal thickness [16,19], which is also consistent with our observation.

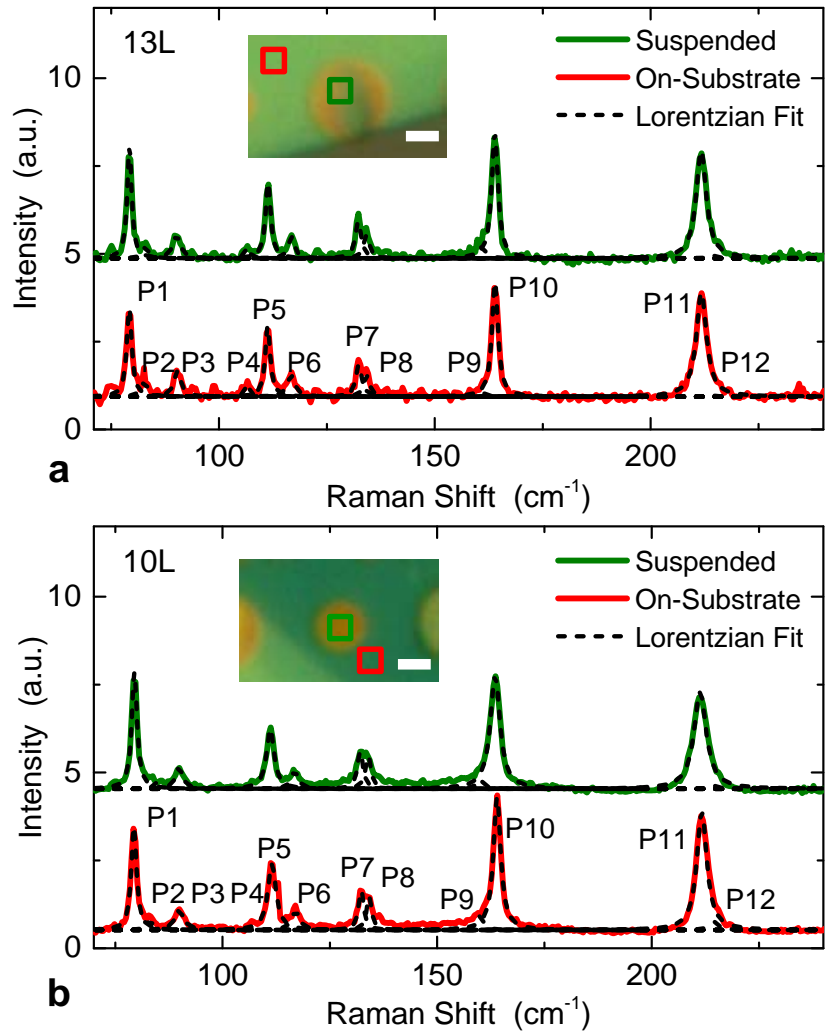

**Figure 3**: Measured Raman spectra of both suspended and on-substrate WTe$_2$ from (a) 13-layer (13L) and (b) 10-layer (10L) samples. Insets show the optical microscope images, with colored box showing location where data (plotted in corresponding colors) are taken. Fitting is shown for data in (a) and (b). Scale bars: 2μm.

We now examine the Raman spectra from both suspended and on-substrate regions of the same WTe$_2$ flakes. Figure 3a shows the Raman spectra from a 13L WTe$_2$ crystal. In this sample, WTe$_2$ flake partially covers the microtrench, making a leaking aperture. Hence, when we locate it in the vacuum chamber, pressure level both inside and outside of cavity quickly equilibrate to vacuum. We find a slight FWHM narrowing in suspended WTe$_2$ Raman modes compared to those from on-substrate WTe$_2$. This may be attributed to reduced coupling to the substrate and minimal damping from air molecules. In contrast, a completely sealed 10L WTe$_2$ diaphragm (see Fig. 3b) which sustains atmosphere pressure inside the cavity when placed in the vacuum chamber, exhibits slightly broader FHWM in all Raman modes from suspended WTe$_2$, than those measured from on-substrate regions on the same flake, suggesting damping from air molecules may play a role in suspended WTe$_2$.

## Properties of Suspended WTe$_2$ via Resonant Measurements

We investigate resonance characteristics of suspended WTe$_2$ structures. Fig. 4a–d show measured resonances from WTe$_2$ circular drumhead resonators of varying dimensions, with resonance frequencies of 8.6–32.1MHz, all in the high frequency (HF) and very high frequency (VHF) bands, and quality ($Q$) factors of 67–249. We summarize resonance frequencies and $Q$ values of all measured devices in Fig. 4e and Fig. 4f.

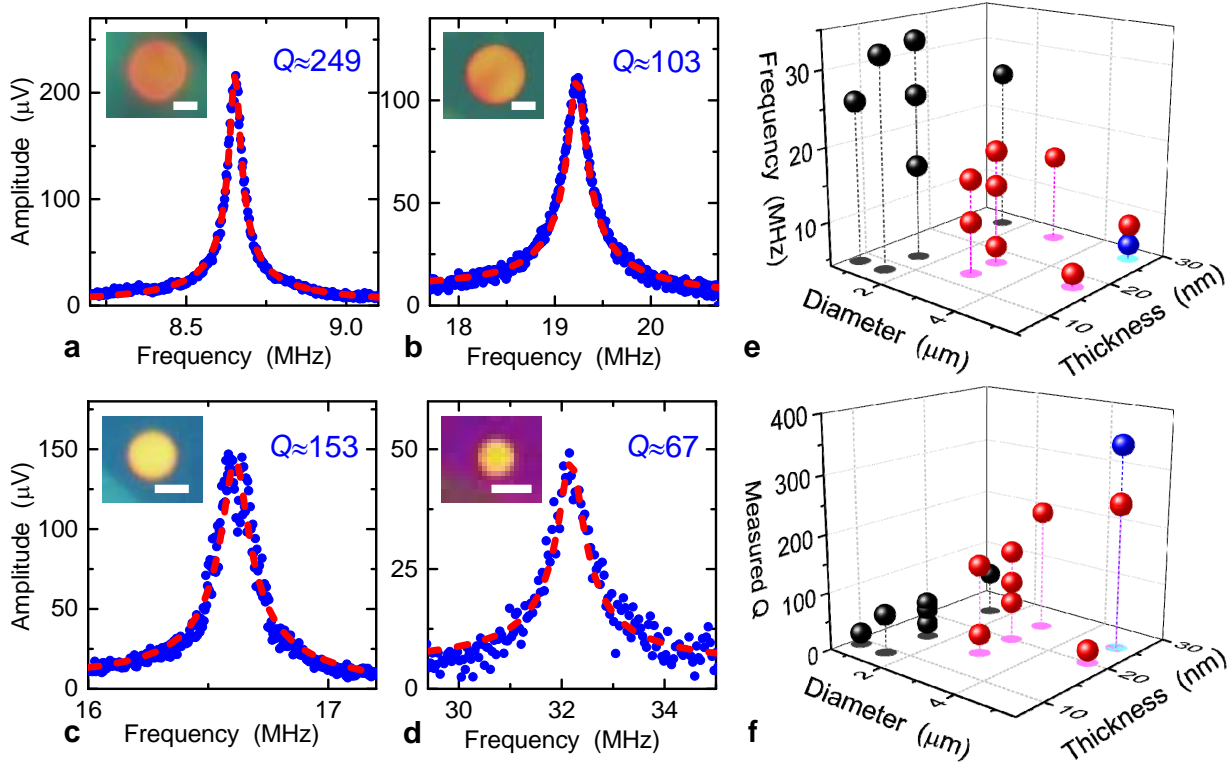

**Figure 4**: Resonance characteristics of suspended WTe$_2$ devices. (a-d) Measured fundamental resonances from devices with dimensions of (a) $d\approx5\mu m$, $t\approx27nm$, (b) $d\approx5\mu m$, $t\approx17nm$, (c) $d\approx3\mu m$, $t\approx13nm$, and (d) $d\approx1.7\mu m$, $t\approx8nm$, respectively. Dashed lines show fits to damped harmonic resonator model (from which $Q$ values are extracted). Insets show optical microscope images of each device. All scale bars: 2μm. (e) Resonance frequencies and (f) quality factors *vs.* device dimensions are summarized for all measured devices (color symbol: black: $d<2\mu m$; red: $d>2\mu m$; blue: partially covered device).

We further perform analytical modeling on frequency scaling and compare with the measurement results. The fundamental-mode resonance frequency $f_0$ of the WTe$_2$ resonators can be expressed as [24,25]:

$$f_0 = \left(\frac{kd}{4\pi}\right)\sqrt{\frac{16D}{\rho_{2D}d^4}\left[\left(\frac{kd}{2}\right)^2 + \frac{\gamma d^2}{2D}\right]}. \qquad (1)$$

Here, $\rho_{2D}$ is the area density ($\rho_{2D}=\rho_{3D}\times t$, where $\rho_{3D}=9430$kg/m$^3$ for WTe$_2$), $k$ a modal parameter (calculated numerically), $\gamma$ the built-in tension or pre-tension, and $D=E_Y t^3/\left[12(1-\nu^2)\right]$ the flexural rigidity ($E_Y$: Young's modulus; $\nu$: Poisson's ratio). Since there is no experimental data for exact Poisson's ratio of WTe$_2$, we have used the value of $\nu=0.16$ predicted by theoretical calculation [26]. In the $\frac{\gamma d^2}{D} \to \infty$ tension-dominant limit (or $\frac{\gamma d^2}{D} \to 0$ modulus-dominant limit), Eq. 1 approaches a membrane (or disk) model (see dashed lines in Fig. 5).

We extract Young's modulus and built-in tension levels of the WTe$_2$ resonators by comparing measurement results with analytical calculations (Eq. 1). Figure 5 plots the frequency scaling of WTe$_2$ resonators using Eq. 1 with diameters of 1.7µm, 3µm, and 5µm, clearly showing the elastic transition from the 'membrane limit' (tension dominating) to the "disk limit" (flexural rigidity dominating) as crystal thickness increases. The resonantly tested devices (with thickness from 8nm to 27nm) operate in the transition regime, where their resonance frequencies are determined by both Young's modulus and the pre-tension. From the data we extract averaged Young's modulus of $E_Y \approx 80$GPa with standard deviation of $\sigma \approx 30$GPa, and built-in pre-tension levels of ~0.05–0.5N/m. These results are consistent with theoretically predicted Young's modulus of WTe$_2$ [27,28]. We note that in this mixed elasticity model, it is more important to find both the tension and $E_Y$ values. In order to determine $E_Y$ more precisely, we shall resort to devices that are completely in the disk regime, and are almost independent of tension. In our experiment, we have not seen any thickness dependence of Young's modulus. Further, we have verified that the uncertainty of Poisson's ratio does not give noticeable deviation to our frequency scaling and Young's modulus estimation. In fact, using different Poisson's ratio values (such as using $\nu=0.3$ to replace $\nu=0.16$) into Eq. 1 only generates negligible differences.

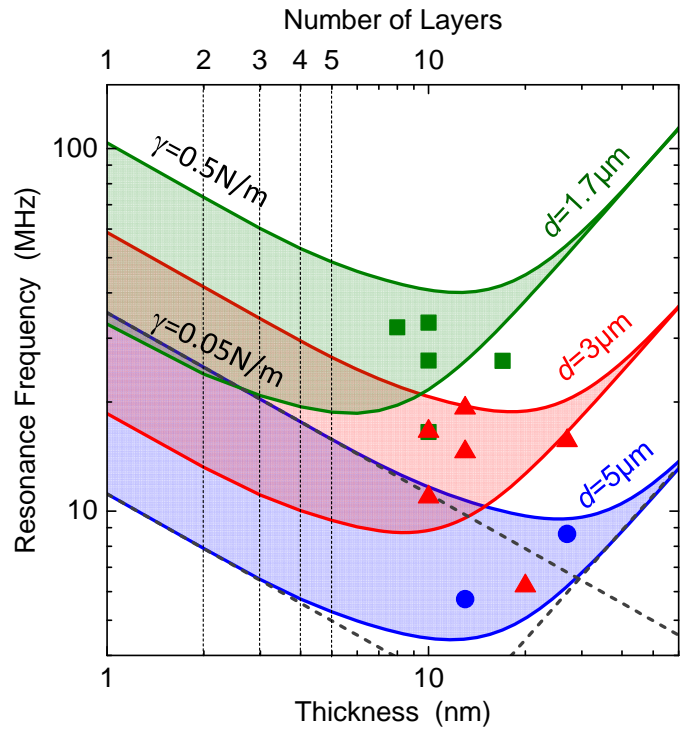

**Figure 5**: Frequency scaling of circular drumhead WTe$_2$ resonators. For each color, upper solid line represents calculated resonance frequency with tension level of $\gamma=0.5$N/m and lower one represents $\gamma=0.05$N/m. The shadowed region indicates the tension levels in between. Dashed lines show membrane and disk limits of $d \approx 5$µm resonators. Squares, triangles, and circles show measured data from devices of $d \approx 1.7$µm, 3µm and 5µm, respectively.

## Conclusions

We have characterized both on-substrate and suspended few-layer $WTe_2$ films by Raman spectroscopy. For the first time, we have observed a total of 12 Raman modes with small FWHM values, in few-layer $WTe_2$ crystals, with new modes and features not yet reported in previous Raman measurements. All the 12 Raman peaks exhibit distinctive thickness dependence (stiffening, softening, or invariant) as crystal thickness decreases from 8L to 1L, which can be used as an effective 'thickness indicator' for $WTe_2$ flakes. These multimode Raman spectra can thus be collectively examined as characteristic fingerprints or signatures for single- and few- to multi-layer $WTe_2$. We have also demonstrated the first HF and VHF $WTe_2$ resonators, and by combining the measured resonance responses with frequency scaling analysis, we extract Young's modulus of these $WTe_2$ flakes ($E_Y \approx 80$ GPa) along with their pre-tension levels (~0.05–0.5N/m). These results shall open new opportunities for nanomechanical $WTe_2$ devices, such as strain-engineered and resonantly tuned magnetoresistance sensors.

**Acknowledgement:** We thank the support from Case School of Engineering, National Academy of Engineering (NAE) Grainger Foundation Frontier of Engineering (FOE) Award (FOE2013-005), National Science Foundation CAREER Award (Grant ECCS-1454570), and CWRU Provost's ACES+ Advance Opportunity Award. Part of the device fabrication was performed at the Cornell NanoScale Science and Technology Facility (CNF), a member of the National Nanotechnology Infrastructure Network (NNIN), supported by the National Science Foundation (Grant ECCS-0335765). Work at Tulane is supported by the DOE under grant DE-SC0014208 and the Louisiana Board of Regents under grant LEQSF (2015-18)-RD-A-23.

# Methods

**Single-Crystal WTe$_2$ Synthesis**

Single crystal WTe$_2$ is grown by a chemical transport method [14]. Stoichiometric Te powder and W powder are put in sealed tube filled with Br$_2$. The tube is placed in a double-zone furnace with temperature gradient of 100°C between 750°C and 650°C. After one week, large single crystal is synthesized. While other TMDCs, such as MoS$_2$ and WSe$_2$ exist in 2H crystal structure, WTe$_2$ exists in a distortion octahedral structure, also known as 1T' structure, under ambient conditions. Thickness of devices is confirmed by atomic force microscopy (AFM) with tapping mode. Figure 2b–e show AFM images and height traces.

**Suspended Device Fabrication**

WTe$_2$ flakes are exfoliated from the bulk WTe$_2$ crystal onto 290nm SiO$_2$ on Si substrate. Suspended devices are fabricated using a dry-transfer method [29]: WTe$_2$ flakes are first exfoliated onto PDMS stamps and then transferred onto pre-patterned microtrenches. Since it is reported that ultrathin WTe$_2$ could degrade quickly in air [16,19], we store devices in vacuum immediately after fabrication.

**Raman Scattering Measurement**

Raman measurements are performed using a customized micro-Raman system that is integrated into an optical interferometric resonance measurement system (Fig. 1c). WTe$_2$ flakes are preserved in a vacuum chamber with optical window. The 532nm laser is focused on the WTe$_2$ flakes in the vacuum chamber using a 50× microscope objective. Typical laser spot size is ~1μm and laser power is limited to below 200μW to avoid excessive laser heating. Raman scattered light from the sample is collected in backscattering geometry and then guided to a spectrometer (Horiba iHR550) with a 2400g/mm grating. Raman signal is recorded using a liquid-nitrogen-cooled CCD. The spectral resolution of our system is ~1cm$^{-1}$. Unlike measurements that are conducted in air, we measure Raman scattering of WTe$_2$ in vacuum ($p$~10mTorr). This vacuum environment can reduce effects from surface adsorbents such as O$_2$ and H$_2$O which may lead to WTe$_2$ crystal degradation.

**Interferometric Resonance Measurement**

We study WTe$_2$ nanomechanical resonances using an ultrasensitive laser interferometry system (Fig. 1c). We photothermally excite resonant motions in suspended WTe$_2$ structures using an amplitude modulated 405nm blue laser. To avoid excessive heating, the blue laser is focused ~5μm away from the devices and power is limited to below 300μW. The modulation depth and frequency of the 405nm laser is controlled by a network analyzer (HP3577A), sweeping from 1MHz to 50MHz. A 633nm red laser with an average power of 600μW is focused onto the center of the WTe$_2$ devices to detect the resonant motions. Typical laser spot sizes are ~5μm and ~1μm for the 405nm and 633nm lasers, respectively. The output signal in the frequency domain is recorded by the same network analyzer.

## References

1  K. S. Novoselov, A. K. Geim, S. V. Morozov, D. Jiang, Y. Zhang, S. V. Dubonos, I. V. Grigorieva and A. A. Firsov, *Science*, 2004, **306**, 666–669.

2  C. Lee, H. Yan, L. E. Brus, T. F. Heinz, J. Hone and S. Ryu, *ACS Nano*, 2010, **4**, 2695–2700.

3  H. Fang, S. Chuang, T. C. Chang, K. Takei, T. Takahashi and A. Javey, *Nano Lett.*, 2012, **12**, 3788–3792.

4  C. Ruppert, O. B. Aslan and T. F. Heinz, *Nano Lett.*, 2014, **14**, 6231–6236.

5  X. Cui, G.-H. Lee, Y. D. Kim, G. Arefe, P. Y. Huang, C. Lee, D. A. Chenet, X. Zhang, L. Wang, F. Ye, F. Pizzocchero, B. S. Jessen, K. Watanabe, T. Taniguchi, D. A. Muller, T. Low, P. Kim and J. Hone, *Nat. Nanotechnol.*, 2015, **10**, 534–540.

6  H. Zeng, J. Dai, W. Yao, D. Xiao and X. Cui, *Nat. Nanotechnol.*, 2012, **7**, 490–493.

7  J. Lee, Z. Wang, K. He, J. Shan and P. X.-L. Feng, *ACS Nano*, 2013, **7**, 6086–6091.

8  K. I. Bolotin, K. J. Sikes, Z. Jiang, M. Klima, G. Fudenberg, J. Hone, P. Kim and H. L. Stormer, *Solid State Commun.*, 2008, **146**, 351–355.

9  X. Zhang, D. Sun, Y. Li, G.-H. Lee, X. Cui, D. Chenet, Y. You, T. F. Heinz, J. Hone, *ACS Appl. Mater. Interfaces*, 2015, **105**, 25923–25929.

10 C. Lee, X. Wei, J. W. Kysar and J. Hone, *Science*, 2008, **321**, 385–388.

11 Q. Wei and X. Peng, *Appl. Phys. Lett.*, 2014, **104**, 251915.

12 H. H. Pérez Garza, E. W. Kievit, G. F. Schneider and U. Staufer, *Nano Lett.*, 2014, **14**, 4107–4113.

13 M. N. Ali, J. Xiong, S. Flynn, J. Tao, Q. D. Gibson, L. M. Schoop, T. Liang, N. Haldolaarachchige, M. Hirschberger, N. P. Ong and R. J. Cava, *Nature*, 2014, **514**, 205–208.

14 P. L. Cai, J. Hu, L. P. He, J. Pan, X. C. Hong, Z. Zhang, J. Zhang, J. Wei, Z. Q. Mao and S. Y. Li, *Phys. Rev. Lett.*, 2015, **115**, 057202.

15 M. K. Jana, A. Singh, D. J. Late, C. R. Rajamathi, K. Biswas, C. Felser, U. V Waghmare and C. N. R. Rao, *J. Phys. Condens. Matter*, 2015, **27**, 285401.

16 Y. Kim, Y. I. Jhon, J. Park, J. H. Kim, S. Lee and Y. M. Jhon, 2015, arXiv:1508.03244 [cond-mat.mes-hall].

17 W. D. Kong, S. F. Wu, P. Richard, C. S. Lian, J. T. Wang, C. L. Yang, Y. G. Shi and H. Ding, *Appl. Phys. Lett.*, 2015, **106**, 081906.

18 Y. Jiang, J. Gao and L. Wang, 2015, arXiv:1501.04898 [cond-mat.mtrl-sci].

19 C.-H. Lee, E. Cruz-Silva, L. Calderin, M. A. T. Nguyen, M. J. Hollander, B. Bersch, T. E. Mallouk and J. A. Robinson, *Sci. Rep.*, 2015, **5**, 10013.

20 K. A. N. Duerloo, Y. Li and E. J. Reed, *Nat. Commun.*, 2014, **5**, 4214.

21 G. Lanzani, *Photophysics of Molecular Materials: From Single Molecules to Single Crystals*, John Wiley & Sons, New York, 2006.

22 H. Terrones, E. Del Corro, S. Feng, J. M. Poumirol, D. Rhodes, D. Smirnov, N. R. Pradhan, Z. Lin, M. A. T. Nguyen, A. L. Elı́as, T. E. Mallouk, L. Balicas, M. A. Pimenta, M. Terrones, *Sci. Rep.*, 2014, **4**, 4215.

23 H. Li, Q. Zhang, C. C. R. Yap, B. K. Tay, T. H. T. Edwin, A. Olivier and D. Baillargeat, *Adv. Funct. Mater.*, 2012, **22**, 1385–1390.

24 H. Suzuki, N. Yamaguchi and H. Izumi., *Acoust. Sci. & Tech.,* 2009, **30**, 348–354.

25 T. Wah, *J. Acoust. Soc. Am.*, 1962, **34**, 275–281.

26 F. Zeng, W.-B. Zhang, B.-Y. Tang, *Chin. Phys. B.*, 2015, **24**, 097103.

27 P. Wagner, V. V Ivanovskaya, M. J. Rayson, P. R. Briddon and C. P. Ewels, *J. Phys. Condens. Matter*, 2013, **25**, 155302.

28 J. Li, N. V, Medhekar and V. B. Shenoy, *J. Phys. Chem. C*, 2013, **117,** 15842–15848.

29 R. Yang, X. Zheng, Z. Wang, C. J. Miller and P. X.-L. Feng, *J. Vac. Sci. Technol. B*, 2014, **32**, 061203.